# A Multirate Numerical Scheme for Large-Scale Vehicle Traffic Simulation

V. V. Kurtc, I. E. Anufriev

*St. Petersburg Polytechnic University, Russia*

# Быстрая схема численного интегрирования с кратными шагами для задачи моделирования автомобильного трафика в масштабах мегаполисов

В. В. Курц, И. Е. Ануфриев

*Санкт-Петербургский государственный политехнический университет, Россия*

***Keywords***: multirate time stepping, stability, citywide traffic simulation, ordinary differential equations.

*Nowadays the city-wide traffic contains hundreds of thousands of vehicles with different scenarios of their behavior. If a microscopic approach is used it leads to solving tremendous systems of ordinary differential equations (ODE) whose components have a wide range of variation rates. The given paper has introduced a multirate numerical scheme with a self-adjusting time stepping strategy. Instead of using a single step size for the whole system we have determined the step size for each component by estimating its own local variation. We also performed the stability analysis for a developed scheme. The presented multirate scheme provides a significant speed-up in processor times compared to the corresponding single-rate one. The use of multiple time steps admits parallel computing.*

***Ключевые слова***: солверы с кратными шагами, устойчивость, моделирование автомобильного трафика в масштабах мегаполисов, обыкновенные дифференциальные уравнения.

*В случае моделирования автомобильного трафика в масштабах крупных городов количество транспортных средств с различными сценариями поведения может достигать десятков тысяч. Если при этом используется микроскопический подход, необходимо решать системы обыкновенных дифференциальных уравнений большой размерности. Скорость изменения величин компонент таких систем обычно лежит в широком диапазоне. В данной статье мы предлагаем схему численного интегрирования с кратными шагами. В отличие от стандартных методов, подразумевающих единый для всех компонент шаг интегрирования, в данном случае для каждой компоненты используется индивидуальный шаг, полученный на основе оценки ошибки численного интегрирования. Для предложенной схемы проведено исследование ее устойчивости. Разработанный численный метод демонстрирует существенное ускорение по сравнению с соответствующим «односкоростным» методом. Использование кратных шагов допускает распараллеливание процесса вычислений.*

# 1. Introduction

Today the number of vehicles amounts hundreds of thousands or even more in case of a city-wide traffic simulation. Moreover, sometimes it is required to simulate vehicle's dynamics with high quality and precision on the one hand and simulations should be in a real-time on the other. In order to achieve these goals a microscopic approach, advanced traffic models and fast numerical methods are preferable. Regarding the city-wide traffic there are a lot of different scenarios of vehicles dynamics such as cruising, accelerating, braking and emergency braking, lane changing and etc. As a result, the speeds of some vehicles are changing much more rapidly than those of the others. Standard single-rate time integration methods for Ordinary Differential Equations (ODEs) work with time steps that are varying in time but remain constant within its components. Consequently, the whole system is forced to take a time step to suit the fastest component which is the smallest time step that can significantly increase the computational time. Multirate time stepping schemes could be efficient to solve such problems. Using such schemes different solution components may be integrated with different time steps.

The given paper is organized as follows. At first, we discuss the work related to multirate methods. Section 2 introduces a microscopic traffic model and a basic numerical integration scheme. We present our multirate approach with an automatic time stepping strategy and discuss the size of a macro time step in Section 3. Then the local integration error for each component is derived and the time step rule is presented. We perform the stability analysis in Section 4. Section 5 contains some numerical test results and the comparison of a multirate method presented with the corresponding single-rate one. The work ends with conclusions and recommendations for our future research.

# 2. Background Material and Preliminaries

## 2.1. Related Work

The first work devoted to multirate numerical methods was conducted and described by Gear and Wells [4] for linear multistep methods. Gunther, Kværnø and Rentrop [5] introduced a multirate scheme which was based on partitioned Runge–Kutta methods with coupling between active (fast) and latent (slow) components performed by the interpolation and extrapolation of state variables. The partitioning into active and latent components is done *a priori* before solving the problem and based on the knowledge of the ODE system to be solved. A related scheme, based on Rosenbrock or ROW methods, was studied by Bartel and Gunther [1]. Kværnø presented some stability results for the simplified versions of these schemes when the systems of two linear equations have one fast and one slow component [8]. An algorithm based on finite elements was proposed by Logg [9; 10].

Multirate schemes for non-stiff problems and explicit methods were examined by Engstler and Lubich [2; 3]. Their strategy uses the extrapolation and implies the automatic partitioning into different levels of slow and fast components during the extrapolation process. Savcenco et al. introduced a self-adjusting multirate time stepping strategy for implicit methods suitable for stiff or mildly stiff ODEs [11]. Hundsdorfer and Savcenco examined the local

V. V. KURTC, I. E. ANUFRIEV. A Multirate Numerical Scheme for Large-Scale Vehicle Traffic Simulation

accuracy, the propagation of interpolation errors and the stability for a simple multirate scheme consisting of the $\theta$-method with one level of a temporal local refinement [6].

### 2.2. Traffic Model and Basic Numerical Scheme

We consider the modification of the IDM (Intelligent Driver Model) [7] as a microscopic traffic model whose acceleration function is as follows

$$\dot{v} = w \cdot a\left(1 - \left(\frac{v}{v^0}\right)^\delta\right) + (1-w) \cdot a\left(1 - \left(\frac{d^*}{h}\right)^2\right). \quad (2.1)$$

All parameters have the same meaning as for the IDM [14]. Additionally, $w$ is a continuous weight function which depends on the distance to the leader $h$ and the parameter $D$. This model is able to reproduce all common traffic modes and phenomena.

$$w = w(h, d^*, D) = \begin{cases} 0, h \in (-\infty, d^*) \\ -2t^3 - 3t^2 + 1, t = (h - d^*)/D - 1, h \in [d^*, d^* + D]. \\ 1, h \in (d^* + D, +\infty) \end{cases} \quad (2.2)$$

Let us consider $N$ vehicles moving one after another. The acceleration function of $i$th vehicle $a^{(i)}$ is defined as (2.1) but each vehicle has its own values of parameters. Finally, we have the ODE system consisting of $2N$ equations with the given initial values for speeds $\{v_{i0}\}_{i=1}^N$ and gaps $\{h_{i0}\}_{i=1}^N$

$$\begin{cases} \dot{v}_1 = a^{(1)}(v_1, h_1) \\ \dot{h}_1 = v_L - v_1 \\ \langle ... \rangle \\ \dot{v}_N = a^{(N)}(v_N, h_N) \\ \dot{h}_N = v_{N-1} - v_N \end{cases} \quad (2.3)$$

$$v_i(0) = v_{i0}, h_i(0) = h_{i0}, i = 1, ..., N.$$

Here $v_i$ and $h_i$ are the speed and the distance to the leader for $i$th vehicle respectively.

In this paper we use the explicit Euler method as the basic numerical integration scheme. For brevity, let us rewrite the system (2.3) in the form of

$$\dot{x}(t) = F(x(t)), x(0) = x_0 \quad (2.4)$$

with the initial value in $x_0 \in \Re^{2N}$. The approximation at the time level $t_n$ is designated by $x_n$. To proceed from $t_{n-1}$ to a new time level $t_n = t_{n-1} + \Delta t$, the method calculates:

$$x_{n+1} = x_n + \Delta t F(x_n). \quad (2.5)$$

## 3. Local Discretization Errors

### 3.1. Multirate Approach with Automatic Time Stepping Strategy

Our multirate numerical scheme is based on a local integration error estimation. Let us consider one macro step from time $t_{n-1}$ to $t_n = t_{n-1} + \Delta T$ with the step size $\Delta T$. Choosing the size of $\Delta T$ will be further discussed in Section 3.2. Within one macro step, for each solution component $x_i$ we consider $k_i$ sequential micro time steps $\Delta t_i = \Delta T / k_i$ according to (2.5). The multiplicity factors $\{k_i\}_{i=1}^N$ are determined to fulfill the accuracy condition:

$$I_i = \|x_n - \widetilde{x}(t_n)\|_\infty < \varepsilon. \qquad (3.1)$$

Here $\widetilde{x}$ is the analytical solution to the problem:

$$\dot{x}(t) = F(x(t)), \, x(t_{n-1}) = x_{n-1}. \qquad (3.2)$$

The maximum norm is used because we need errors below the tolerance for all components.

Here it is worth noting that with our multirate approach component values may be needed in the course of the macro time step, but those are not calculated at certain points of a current time interval. In our case we consider them equal to the values at the beginning of this current macro step and take this fact into account when calculating a local integration error $I_i$. The solutions are synchronized when each macro time step is completed.

### 3.2. Determining the Size of Macro Time Step

The macro time step $\Delta T$ is defined *a priori* and considered to be constant during the whole integration process in our multirate scheme. The value of $\Delta T$ is determined as follows. On the one hand, one should choose it as high as possible to reduce the number of synchronization operations as they are time-consuming to some extent. On the other hand, $\Delta T$ cannot be better than the driver reaction time, since within one macro time step a driver knows speeds and positions of all other vehicles only at the beginning of this macro step and a lot of nontrivial things can happen like the emergency braking of the leader or instant lane changes of neighboring vehicles. That is why in our algorithm $\Delta T = 0.5s$ that is equal to the average value of the driver reaction time.

### 3.3. Step Size Rule and Automatic Time Stepping Strategy

Let us consider the *i*th solution component of (2.4) and determine the number of sequential micro time steps $k_i$ and the corresponding micro time step value $\Delta t_i = \Delta T / k_i$ which ensure the fulfillment of the accuracy condition (3.1). To proceed from $t_m$ to next time level $t_m + \Delta T$, the method calculates:

$$\begin{cases} x_i^{(m+1)} = x_i^{(m)} + \Delta t_i \cdot f_i\left(x_1^{(m)}, x_2^{(m)}, \ldots, x_i^{(m)}, \ldots, x_n^{(m)}\right) \\ x_i^{(m+2)} = x_i^{(m+1)} + \Delta t_i \cdot f_i\left(x_1^{(m)}, x_2^{(m)}, \ldots, x_i^{(m+1)}, \ldots, x_n^{(m)}\right) \\ \ldots \\ x_i^{(m+k_i)} = x_i^{(m+k_i-1)} + \Delta t_i \cdot f_i\left(x_1^{(m)}, x_2^{(m)}, \ldots, x_i^{(m+k_i-1)}, \ldots, x_n^{(m)}\right). \end{cases} \qquad (3.3)$$

Here $f_i = (F)_i$. By simple transformations of (3.3) we obtain the relation between the solutions $x_i^{(m+k)}$ and $x_i^{(m)}$ in adjacent macro time layers:

$$x_i^{(m+k_i)} = x_i^{(m)} + \Delta t_i \sum_{j=0}^{k_i - 1} f_i\left(x_1^{(m)}, x_2^{(m)}, \ldots, x_i^{(m+j)}, \ldots, x_n^{(m)}\right). \quad (3.4)$$

Now we will gradually calculate $x_i^{(m+j)}$, $j = 2, \ldots, k_i$ expressing it in terms of the solution $x^{(m)}$ at the start of this current macro time step. For the $x_i^{(m+2)}$ we have

$$x_i^{(m+2)} = x_i^{(m)} + \Delta t_i f_i\left(x^{(m)}\right) + \Delta t_i f_i\left(x_1^{(m)}, \ldots, x_i^{(m)} + \Delta t_i f_i\left(x^{(m)}\right), \ldots, x_n^{(m)}\right). \quad (3.5)$$

Expanding the third term in Tailor series in the neighborhood of $x^{(m)}$ we obtain

$$x_i^{(m+2)} = x_i^{(m)} + 2\Delta t_i f_i\left(x^{(m)}\right) + \Delta t_i^2 f_i\left(x^{(m)}\right)\frac{\partial f_i}{\partial x_i}\left(x^{(m)}\right) + O\left(\Delta t_i^3\right). \quad (3.6)$$

Performing the same procedure for $x_i^{(m+3)}$ we get

$$x_i^{(m+3)} = x_i^{(m)} + 3\Delta t_i f_i\left(x^{(m)}\right) + 3\Delta t_i^2 f_i\left(x^{(m)}\right)\frac{\partial f_i}{\partial x_i}\left(x^{(m)}\right) + O\left(\Delta t_i^3\right).$$

Finally, after $k_i$ sequential steps we obtain

$$x_i^{(m+k)} = x_i^{(m)} + k_i \Delta t_i f_i\left(x^{(m)}\right) + C_{k_i}^2 \Delta t_i^2 f_i\left(x^{(m)}\right)\frac{\partial f_i}{\partial x_i}\left(x^{(m)}\right) + O\left(\Delta t_i^3\right). \quad (3.7)$$

Here $C_{k_i}^2 = \binom{k_i}{2} = \frac{k_i(k_i - 1)}{2}$.

Now let us determine the analytical solution $\tilde{x}_i$ to the problem (3.2) with the initial value $\tilde{x}(t_m) = x^{(m)}$ at the end of the current macro time step $t_m + \Delta T$

$$\tilde{x}_i(t_m + T) = \tilde{x}(t_m) + \Delta T f_i\left(x^{(m)}\right) + \frac{\Delta T^2}{2} \sum_{j=1}^{n} \frac{\partial f_i}{\partial x_j}\left(x^{(m)}\right) f_j\left(x^{(m)}\right) + O\left(\Delta T^3\right). \quad (3.8)$$

Subtracting (3.7) from (3.8) we obtain the local integration error after one macro time step $(t_m \to t_m + \Delta T)$

$$e_{i,k_i} = \left|\frac{\Delta T^2}{2}\left(\sum_{\substack{j=1 \\ j \neq i}}^{n} \frac{\partial f_i}{\partial x_j}\left(x^{(m)}\right) f_j\left(x^{(m)}\right) + \frac{1}{k_i}\frac{\partial f_i}{\partial x_i}\left(x^{(m)}\right) f_i\left(x^{(m)}\right)\right) + O\left(\Delta T^3\right)\right|. \quad (3.9)$$

Discarding $O(\Delta T^3)$ we have the estimation for the local integration error for $i$th solution component:

$$est_{i,k_i} = \frac{\Delta T^2}{2}\left|\sum_{\substack{j=1 \\ j \neq i}}^{n} \frac{\partial f_i}{\partial x_j}\left(x^{(m)}\right) f_j\left(x^{(m)}\right) + \frac{1}{k_i}\frac{\partial f_i}{\partial x_i}\left(x^{(m)}\right) f_i\left(x^{(m)}\right)\right|. \quad (3.10)$$

If the number of micro steps within one macro step is equal to 1, that is $k_i = 1$, the estimation (3.10) is transformed to the classical and well-known expression for the single-rate explicit Euler method [12].

For the user-specified accuracy $\varepsilon$ for each component $x_i$, using the error estimation (3.10) one can determine the minimum value of $k_i$ which ensures the solution with the tolerance below $\varepsilon$

$$\min_{k_i \in \mathbb{N}} k_i : err_{i,k_i} < \varepsilon. \quad (3.11)$$

As applied to the traffic flow ODE system (2.3), the local error estimation for the speed of $i$th vehicle will be as follows:

$$est\_v_{i,k_i} = \frac{\Delta T^2}{2k_i} \left| a_v^{(i)} a^{(i)} + a_h^{(i)}(v_{i-1} - v_i) \right|. \quad (3.12)$$

Here $a_v^{(i)}$ and $a_h^{(i)}$ are partial derivatives of the acceleration function $a^{(i)}$ (2.1). In this case we perform the error control only for speed values. Finally, the micro time step $\Delta t_i$ for $i$th vehicle, which ensures that the magnitude of the estimated local error $est\_v_{i,k_i}$ is within the tolerance $\varepsilon_v$, is as follows:

$$\Delta t_i = \frac{\Delta T}{\left\lceil \frac{\Delta T^2}{2\varepsilon_v} \left| a_v^{(i)} a^{(i)} + a_h^{(i)}(v_{i-1} - v_i) \right| \right\rceil}. \quad (3.13)$$

Here $\lceil \cdot \rceil$ is the rounding function towards the positive infinity. When the flow is homogeneous, i.e. the denominator of (3.13) is equal to zero, the time step $\Delta t_i$ is set to the macro time step $\Delta T$.

## 4. Stability

The related stability results for multirate methods can be found in [5; 8; 13; 15]. Here we perform the stability analysis of our multirate scheme in the application to the ODEs system (2.3) which describes the traffic flow dynamics.

Let us consider one macro step in accordance with (3.3) for the whole system (2.3)

$$\begin{cases} x_1^{(m+1)} = x_1^{(m)} + \Delta t_1 \cdot f_1\left(x_1^{(m)}, x_2^{(m)}, \ldots, x_n^{(m)}\right) \\ \ldots \\ x_1^{(m+k_1)} = x_1^{(m+k_1-1)} + \Delta t_1 \cdot f_1\left(x_1^{(m+k_1-1)}, x_2^{(m)}, \ldots, x_n^{(m)}\right) \\ x_2^{(m+1)} = x_2^{(m)} + \Delta t_2 \cdot f_2\left(x_1^{(m)}, x_2^{(m)}, \ldots, x_n^{(m)}\right) \\ \ldots \\ x_2^{(m+k_2)} = x_2^{(m+k_2-1)} + \Delta t_2 \cdot f_2\left(x_1^{(m)}, x_2^{(m+k_2-1)}, \ldots, x_n^{(m)}\right). \\ \ldots \end{cases} \quad (4.1)$$

The goal is to define multiplicity factors $k_i$ and the corresponding micro time step $\Delta t_i = \dfrac{T}{k_i}$ for each vehicle at the beginning of the macro time step which ensure the stability of our multirate scheme. Linearizing (2.3) in the neighborhood of $x^{(m)}$ we can write

$$x^{(m+k)} = R(J, \Delta T, k_1, ..., k_n) x^{(m)}. \qquad (4.2)$$

Here $R(J, \Delta T, k_1, ..., k_n)$ is the transition matrix. The Jacobian matrix $J$ of (2.3) has a block-like structure:

$$J = \begin{bmatrix} a_v^{(1)} & a_h^{(1)} & 0 & 0 & 0 & 0 & ... \\ -1 & 0 & 0 & 0 & 0 & 0 & ... \\ 0 & 0 & a_v^{(2)} & a_h^{(2)} & 0 & 0 & ... \\ +1 & 0 & -1 & 0 & 0 & 0 & ... \\ 0 & 0 & 0 & 0 & a_v^{(3)} & a_h^{(3)} & ... \\ 0 & 0 & +1 & 0 & -1 & 0 & ... \\ ... & ... & ... & ... & ... & ... & ... \end{bmatrix}. \qquad (4.3)$$

The method (4.1) is stable if and only if the eigenvalues of $R$ are all within the unit disk

$$R = \begin{bmatrix} r_1^{k_1} & \dfrac{r_1^{k_1}-1}{r_1-1} a_h^{(1)} \dfrac{\Delta T}{k_1} & 0 & 0 & ... \\ -\Delta T & 1 & 0 & 0 & ... \\ 0 & 0 & r_3^{k_3} & \dfrac{r_3^{k_3}-1}{r_3-1} a_h^{(2)} \dfrac{\Delta T}{k_3} & ... \\ +\Delta T & 0 & -\Delta T & 1 & ... \\ ... & ... & ... & ... & ... \end{bmatrix}$$

with $r_{2i-1} = \left(1 + a_v^{(i)} \dfrac{\Delta T}{k_i}\right)$. Due to the block structure of $R$ it is possible to find all its eigenvalues and obtain the stability conditions for our multirate scheme (4.1)

$$\left| 1 + r_{2i-1}^{k_{2i-1}} \pm \sqrt{\left(1 + r_{2i-1}^{k_{2i-1}}\right)^2 - 4 \dfrac{r_{2i-1}^{k_{2i-1}}-1}{r_{2i-1}-1} a_h^{(i)} \dfrac{\Delta T^2}{k_{2i-1}}} \right| < 2, i = 1, ..., n. \quad (4.4)$$

Numerical experiments show that if we consider only the time step rule (3.13) the stability condition (4.4) is violated for specific vehicles at certain times. The main goal is to define $\Delta t_i (i = 1, ..., n)$ in such way that conditions (3.13) and (4.4) are satisfied.

## 5. Numerical Experiments

Let us consider one vehicle following its leader. The initial conditions for (2.3) are $v_1 = 0$ m/s, $h_1 = 100$ m, the speed tolerance is $\varepsilon_v = 0.1$. We offer a certain scenario for the leader and run simulations. Fig. 1 shows that the estimation time step rule (3.13) works correctly, that is, micro time step values provide the solution with the required accuracy $\varepsilon_v = 0.1$.

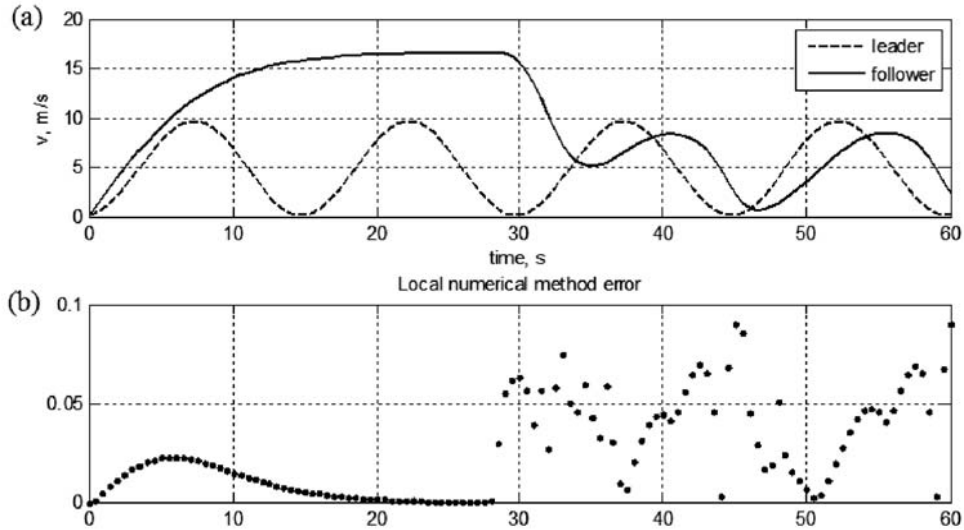

Fig. 1. (a) Velocity time history of leader (dashed) and its follower (solid).
(b) Numerical integration error at each macro time step

We consider the solution obtained by the Forward Euler method with time step $\Delta t = 0.005$ s as the exact solution to calculate numerical integration errors. Some other numerical experiments have been performed and they have demonstrated the correctness of (3.13).

Figure 2 compares two numerical integrating methods – the single-rate Forward Euler method with variable time steps and our multirate scheme. The number of simulated vehicles is 1000. The macro time step for a multirate method is 0.5 s. A time simulation period is 100 s. We consider the following stages for the first method at each time step:

1. To calculate the time step for each vehicle $\{\Delta t_i\}_{i=1}^n$ on the basis of the classical *a priori* error estimate [12] and the solution accuracy $\varepsilon_v$.

2. To determine the time step $\Delta t = min\{\Delta t_i\}_{i=1}^n$.

3. To perform one step for (2.3) with the time step $\Delta t$.

These results demonstrate that the multirate approach prevails over the single-rate one and provides a reasonable speed-up. Moreover, the speed tolerance $\varepsilon_v$ has an impact on the performance of methods. Fig. 2, (a) shows that for $\varepsilon_v = 0.1$ the multirate method is 2 times faster than the forward Euler method with variable time steps. For $\varepsilon_v = 0.5$ the former is 3.3 times faster than the latter one (see Fig. 2, (b)).

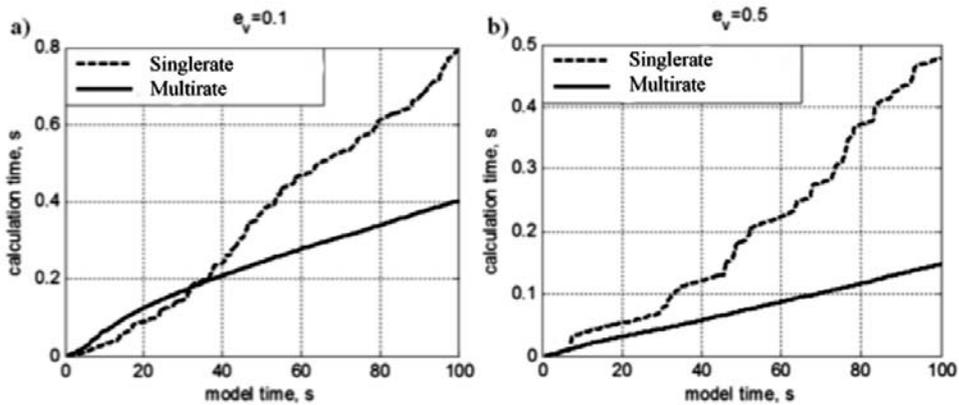

Fig. 2. Calculation time versus model time – multirate method (solid) and single-rate method (dashed). (a) $\varepsilon_v = 0.1$  (b) $\varepsilon_v = 0.5$

## 6. Conclusions

In this paper we have presented the multirate numerical scheme with an automatic time stepping strategy for ODEs. The time step for a particular system component is determined on the basis of the estimation of the local integration error. In this way, we have performed several sequential micro steps within one macro time step for each component. This scheme has been used to integrate the ODEs system describing the traffic flow dynamics. We have used the modified Intelligent Driver Model as a microscopic traffic model. Numerical experiments confirm that the local integration error after every macro integration step does not exceed *a priori* prescribed solution accuracy. Furthermore, numerical results demonstrate that the efficiency of numerical integration methods can be significantly improved by using the large time steps for slow components and the small steps for fast ones without sacrificing the solution accuracy.

We have performed the stability analysis for the presented multirate scheme. Due to the specific structure of the ODEs system solved, it is possible to obtain the stability condition analytically. This condition is obtained implicitly, that is, one cannot calculate directly the micro time step value. However, it is implied that, initially, the micro time step is determined on the basis of the time step rule and then the stability condition should be checked.

It is worth mentioning that the time step rule and the stability condition are obtained in general for any microscopic car-following model whose acceleration function depends on a vehicle speed and the distance to its leader.

As the basic numerical integration method the forward Euler method has been used. We have not performed the data interpolation and/or extrapolation within the macro time step limits in our approach,. Higher-order methods and different interpolation formulas are going to be investigated.

Finally, our multirate approach admits parallel computing. Each micro time step value is multiple to the macro one and, as a result, we have the finite set of micro time steps. Groups are formed, each of which contains equations with the same micro step. The calculations for each are independent and can be performed in parallel. We consider this issue to be promising and it will speed up the calculations even more than those we have already conducted.

**Valentina V. Kurtc**
M.Sc.
St. Petersburg Polytechnic University, postgraduate student.
E-mail: kurtsvv@gmail.com
**Igor E. Anufriev**
Ph.D. (Physical and Mathematical Sciences)
St. Petersburg Polytechnic University, associate professor.
E-mail: igevan@mail.ru